\documentclass[10pt]{article}
\usepackage{amsmath,amssymb,bbm}
\usepackage{amsthm}

\def \FF {\mathfrak{F}}
\def \FF {\mathfrak{F}}
\def \BB {\mathfrak{B}}

\def \DD {\mathcal{D}}

\def \HH {\mathcal{H}}

\def \PPP {\mathbb{P}}

\def \equals {\ = \ }

\def \Tr {\mathrm{Tr}}

\def \bn {{\mathfrak n}}

\newtheorem{proposition}{Proposition}

\newtheorem{definition}{Definition}

\newtheorem*{proposition*}{Proposition}

\author{Alex D. Gottlieb and Norbert J. Mauser}

\title{Correlation in fermion or boson systems as the minimum of entropy relative to all free states}

\begin{document}

\maketitle

\begin{abstract}

In the context of many-fermion systems, ``correlation" refers to the inadequacy of an independent-particle model.   Using ``free" states as archetypes of our independent-particle model, we have proposed a measure of correlation that we called ``nonfreeness" [Int. J. Quant. Inf. {\bf 5}, 815 (2007)].  The nonfreeness of a many-fermion state was defined to be its entropy relative to the unique free state with the same $1$-matrix.

In this article, we prove that the nonfreeness of a state is the minimum of its entropy relative to {\it all} free states.  We also extend the definition of nonfreeness to many-boson states and discuss a couple of examples.
\end{abstract}


\section{Introduction}

We wish to advance one clearly defined measure of correlation for many-fermion or many-boson states.  
To distinguish our notion of correlation from others that have been discussed in the literature \cite{correlation}, we will presently write of ``free" rather than ``uncorrelated" states, and of ``nonfreeness" rather than ``correlation."  
Also, when we write of ``particles" we refer exclusively, if ambiguously, to fermions or bosons.

We understand ``correlation" in the sense of its usage in quantum chemistry or condensed matter physics, where it connotes a shortfall of an independent-particle model.
In quantum chemistry, the customary independent-particle model for describing a many-electron ground state wave function is the Hartree-Fock approximation, the Slater determinant of lowest energy.   For applications in condensed matter physics, the generalized Hartree Fock theory of Bogoliubov or Valatin is a more flexible independent-particle model, admitting a broader class of uncorrelated states called quasi-free states
\cite{BachLiebSolovej, KrausCirac}.   These are the states we deem uncorrelated.  

We regard a many-particle state to be free of correlation if it has the form of a grand-canonical equilibrium state of non-interacting particles.  
The particles may actually be interacting, and the system need not be in an equilibrium state, but  as long as the state of the system {\it has the form} of an equilibrium state of an open system of non-interacting particles, 
we would say that it is free from correlation and call it ``free."

Regarding free states as the least correlated states, we can quantify the amount of correlation that a given state possesses by comparing it to free states.   
In \cite{GottliebMauser2005} and \cite{GottliebMauser2007} we have introduced such measures of fermion correlation, comparing the many-fermion state of interest to the unique free state with the same $1$-matrix. 
To quantify how much a given state resembles the corresponding free state, we used two different functionals, 
i.e., the negative logarithm of fidelity \cite{GottliebMauser2005} and the relative entropy \cite{GottliebMauser2007}.    These are both quantum R\'enyi divergences \cite{Mueller-Lennert...,WildeWinterYang}, the former with parameter $\alpha = \tfrac12$ and the latter with $\alpha = 1$.

In \cite{GottliebMauser2007} we defined the ``nonfreeness" of a many-fermion state to be the entropy of that state relative to the free state with the same $1$-matrix.  
In this article, we shall prove that the nonfreeness of a state is the minimum of the entropy of that state relative to {\it all}  free states.  

We proceed to elaborate upon nonfreeness and its properties, and to state our main result. 

The nonfreeness of a many-particle state equals zero if and only if the state is free, otherwise it is positive.  A state's nonfreeness is a function of its natural occupation numbers and the von Neumann entropy of its density operator.  

A pure $n$-particle state, which can be represented by a normalized wave function $\Psi(x_1,x_2,\ldots,x_n)$ of the appropriate symmetry, has a density operator whose von Neumann entropy is $0$, and its nonfreeness is just a function of its natural occupation numbers $n_i$.  
The nonfreeness of a $n$-fermion state is 
\begin{equation}
\label{nonfreeness fermions pure} 
           - \sum_i  n_i\log(n_i) - \sum_i  (1-n_i)\log(1-n_i) 
\end{equation}
and the nonfreeness of a pure $n$-boson state is  
\begin{equation}
\label{nonfreeness bosons pure} 
           \sum_i  (1+n_i)\log(1+n_i) - \sum_i  n_i\log(n_i) \ . 
\end{equation}
In these formulas, and in all others below, $0\log(0)$ is to be evaluated as $0$.  

The nonfreeness (\ref{nonfreeness fermions pure}) of a pure $n$-fermion states is exactly the same as its ``particle-hole symmetric correlation entropy" that has been introduced in \cite{Gori-GiorgiZiesche} and applied in \cite{Bendazzoli...}.  The correlation entropy (\ref{nonfreeness fermions pure}) of an $n$-fermion wave function $\Psi$ equals $0$ if and only if $\Psi$ is a Slater determinant, in which case $n$ of the natural occupation numbers equal $1$ and the rest equal $0$.

The nonfreeness  (\ref{nonfreeness bosons pure}) of a pure $n$-boson state equals $0$ only if all $n_i=0$, which is the case only for the vacuum state.  A pure $n$-boson state with $n > 0$ is never free.  The freest pure $n$-boson states are the ``atomic coherent states" \cite{Arecchi...} of the form 
\[
    \Psi(x_1,x_2,\ldots,x_n) \ = \ \psi(x_1)\psi(x_2)\cdots \psi(x_n)\ , 
\]
where the occupation of the natural mode $\psi$ is $n$, and the rest of the natural occupation numbers are $0$.

More generally, nonfreeness is defined for {\it mixed} states, i.e., states that are represented by density operators on the many-particle Fock space.  
This may be useful even when one is mainly interested in pure states of a many-particle system, because a {\it subsystem} of that many-particle system, consisting of the particles that occupy a given subset of available modes, is generically in a mixed state.
For example, consider a system of $n$ fermions on a lattice.  The fermions that occupy a given site, or block of contiguous sites, constitute a subsystem that is typically in a mixed state. 
The state of the fermions at a given site or in a block of sites, especially the von Neumann entropy of that state, may reflect physical properties \cite{Zanardi,ZanardiWang} such as quantum phase transitions \cite{VidalLatorreRicoKitaev,GuDengLiLin,LarssonJohannesson}.   
Nonfreeness of single-site subsystems of fermion lattice systems was studied in \cite{MauserHeld}, following a similar study of correlations in (LDA+DMFT) tight-binding models of transition metal oxides \cite{Byczuk,ByczukErrata}.

We restrict our attention to the class of states that are represented by density operators $\Delta$ on the fermion or boson Fock space that (i) commute with the number operator $\hat{n}$, and (ii) have finite average particle number, i.e., such that $\Tr(\hat{n} \Delta ) < \infty$.   Let $\DD$ denote, somewhat ambigously, the class of density operators on the fermion Fock space, or on the boson Fock space, that satisfy conditions (i) and (ii) above.  Note that the pure $n$-particle states are contained in $\DD$, for when $\Psi$ represents an $n$-particle state, the corresponding density operator $\Delta = |\Psi\rangle\!\langle \Psi|$  commutes with $\hat{n}$ and has $\Tr(\hat{n} \Delta ) =n$.      
   
Let $\Delta \in \DD$ be a density operator on the fermion or boson Fock space, representing a many-particle state with natural occupation numbers $n_i$.   The nonfreeness of that state is then 
\begin{eqnarray}
\label{nonfreeness fermions} 
   \hbox{fermions:}  \quad &         - \sum\limits_i  n_i\log(n_i) - \sum\limits_i  (1-n_i)\log(1-n_i) + \Tr(\Delta\log \Delta) \qquad \\
\label{nonfreeness bosons} 
   \hbox{bosons:} \quad &        - \sum\limits_i  n_i\log(n_i)  + \sum\limits_i  (1+n_i)\log(1+n_i)   + \Tr(\Delta\log \Delta)  . \qquad
\end{eqnarray}

The last term in formulas (\ref{nonfreeness fermions}) and (\ref{nonfreeness bosons}) is the negative of the von Neumann entropy of $\Delta$.  
Thus the nonfreeness of $\Delta$ is given by (\ref{nonfreeness fermions}) or (\ref{nonfreeness bosons}) minus the von Neumann entropy of $\Delta$. 
Subtracting the von Neumann entropy of $\Delta$ has 
the effect that the nonfreeness of the state $\Delta$ is at least as great as that of any of its ``substates."  
For example, in a system of itinerant fermions on a lattice, the fermions that occupy a certain site or block of sites constitute an open subsytem that is typically in a mixed state, even when the state of the whole lattice is pure.   The nonfreeness of this mixed state is less than or equal to the nonfreeness of the whole state.   In particular,  all substates of a free state are free.  

This ``monotonicity property" of nonfreeness is not obvious from formulas (\ref{nonfreeness fermions}) or (\ref{nonfreeness bosons}).  It is related to the folllowing, more fundamental, interpretation of nonfreeness as a relative entropy:

The entropy of a density operator $\Delta$  {\it relative to} a density operator $\Gamma$ is  
$$ S(\Delta | \Gamma) \equals -\Tr(\Delta \log \Gamma) + \Tr(\Delta \log\Delta).$$ 
   In \cite{GottliebMauser2007}, the nonfreeness of a density operator $\Delta$ on the many-fermion Fock space was defined to be the entropy of $\Delta$ relative to the unique free state with the same $1$-matrix as $\Delta$, and it was shown there that this relative entropy is given by formula (\ref{nonfreeness fermions}) above.  In this article, we shall define nonfreeness for bosonic states analogously, and we shall show that it is given by formula (\ref{nonfreeness bosons}) above.
   
   Our main result is the following:

\begin{proposition*}
Suppose that $\Delta \in \DD$ is a density operator on the fermion or boson Fock space.  
Let $\Gamma_{\Delta}$ denote the  unique free state that has the same $1$-matrix as $\Delta$.  Then the minimum value of $S(\Delta|\Gamma _{free})$ over all free density matrices is attained by $\Gamma_{\Delta}$, that is, 
$$  S(\Delta|\Gamma_{\Delta}) \ = \ \min \big\{S(\Delta|\Gamma _{free}) : \ \Gamma _{free} \ \mathrm{is}\ \mathrm{ free} \big\} . $$  
\end{proposition*}

In light of this fact, one may view $\Gamma_{\Delta}$ as a kind of optimal independent-particle approximation of $\Delta$, like the ``best-density" and ``best-overlap" Slater determinant approximations discussed in  \cite{KutzelniggSmith}.   The nonfreeness $S(\Delta|\Gamma_{\Delta})$ quantifies how much $\Delta$ differs from its optimal independent-particle approximation.   

Nonfreeness, as the minimum of entropy relative to all free states, is reminiscent of the ``relative entropy of entanglement"  \cite{Vedral,Vedral...}, which is the minimum of entropy relative to all ``separable" states.  However, we must emphasize that we do not intend nonfreeness to be some sort of entanglement measure for indistinguishable particles \cite{emphasize}.

The rest of this article is organized as follows.  
Free density operators are defined in Section~\ref{free states section}, $1$-matrices are discussed in Section~\ref{$1$-matrix section}, and nonfreeness is defined in Section~\ref{nonfreeness definition section}.   The fact that  $\Gamma_{\Delta}$ minimizes $S(\Delta|\Gamma)$ is proved in Section~\ref{the proof section}, where formulas (\ref{nonfreeness fermions}) and (\ref{nonfreeness bosons}) are justified as well.  Two illustrations of nonfreeness are given in Section~\ref{examples section}.  Section~\ref{random substates} concerns random pure states of $n$-fermions in $m$ orbitals, and Section~\ref{ideal canonical bosons} concerns the canonical ensemble for an ideal $n$-boson gas.   After the conclusion, Section~\ref{conclusion}, there is an appendix that explains how we computed values of nonfreeness for the examples in Section~\ref{ideal canonical bosons}

\section{Free states and the concept of nonfreeness}

\subsection{Free states}
\label{free states section}

Physically, free states have the form of Gibbs grand canonical ensembles of {\it non-interacting} fermions or bosons.  We shall briefly review the well known \cite{Pathria} formalism here, in order to motivate Definitions~\ref{free bosons definition} and \ref{free fermions definition} below.

Given a reference system of one-particle ``modes" (often called ``orbitals" in the fermion context) one can describe certain configurations of indistinguishable particles by specifying the number of particles in each mode.  One cannot say {\it which} particle is in a certain mode, for the particles are indistinguishable; one can only say how many particles ``occupy" that mode.   To specify the occupation numbers, that is, how many particles are in each mode, we shall use ``occupation lists" 
\[
     \bn = \big(\bn(1),\bn(2),\bn(3),\ldots \big).
\]
The whole occupation list is denoted $\bn$, and $\bn(i)$ denotes the number of particles in the $i$'th mode.  We will be considering only finite configurations: the total number $\sum_i \bn(i)$ of particles in a configuration is assumed to be finite.  In a configuration of fermions, no mode may be occupied by more than one particle, and $\bn(i)$ is either $0$ or $1$.   In a configuration of bosons, however, the modes can be occupied by any number $\bn(i) \ge 0$ of particles.   
We will denote the sets possible occupation lists for fermions and bosons by $\FF$ and  $\BB$, respectively.

This way of of indexing many-particle configurations depends on the reference system of $1$-particle modes.  In the Hilbert space formalism, the reference system of modes is given by an ordered orthonormal basis of the $1$-particle Hilbert space $\HH$.  The finite configurations, as indexed by occupation lists $\bn$ with $\sum_i \bn(i)<\infty$, correspond to orthonormal vectors $|\bn\rangle$  in a Hilbert space called the Fock space over $\HH$.   Superpositions of states $|\bn\rangle$ are also in the Fock space, and may also represent quantum states of the many-fermion or many-boson system.  The vectors $|\bn\rangle$ constitute an orthonormal basis of the Fock space.  A basis of the fermion Fock space of the form $\big\{ |\bn\rangle :  \bn \in \FF \big\}$, or a basis of the boson Fock space of the form $\big\{ |\bn\rangle: \bn \in \BB \big\}$, is called a ``Fock" basis.  It must be borne in mind that a Fock basis is defined with reference to an ordered orthonormal basis of the $1$-particle space.
Changing the reference basis for $\HH$ induces a change of the Fock basis.

Now, for a system of {\it non-interacting} indistinguishable particles, where the particles are independently subject to the same $1$-particle Hamiltonian $H$, a useful reference basis of $\HH$ is given by the eigenvectors of $H$.  Each mode is an eigenvector of $H$, and the corresponding eigenvalue of $H$ is the energy of each particle occupying that mode.  
Thus, if $f_i$ is an eigenvector of $H$ with $Hf_i = \epsilon_i f_i$, then the energy of $\bn(i)$ non-interacting particles in that mode is 
$\bn(i)\epsilon_i$.  The total energy of the configuration $\bn$ is $\sum_i \bn(i)\epsilon_i$.  The grand canonical ensemble for an open system of such non-interacting particles, in equilibrium at temperature $T$ and chemical potential $\mu$, has partition function
\begin{equation}
\label{partition}
      Z \equals  \sum_{\bn} e^{ -\beta \sum_i \bn(i)  (\epsilon_i - \mu)} \ ,
\end{equation}
where $\beta = 1/k_BT$.  We assume that $Z < \infty$.  In the boson case, this requires the chemical potential $\mu$ to be strictly less than all energies $\epsilon_i$.  

The equilibrium state is represented by the following density operator on the Fock space:
\begin{equation}
\label{Gibbs}
    \Gamma \ = \  \frac{1}{Z}   \sum_{\bn} e^{-\beta \sum_i \bn(i) (\epsilon_i - \mu)} |\bn \rangle\!\langle \bn |   
\end{equation}
where $  |\bn \rangle\!\langle \bn | $ denotes the orthogonal projector onto the span of the Fock space vector $|\bn \rangle$.  
Note that the summations over $\bn$ in (\ref{partition}) and (\ref{Gibbs}) are over different sets of occupation lists $\bn$ in the fermion and boson cases; in the fermion case the summation is restricted to configurations $\bn \in \FF \subsetneq \BB$.

Setting $z_i = \exp( -\beta(\epsilon_i - \mu))$, the partition function (\ref{partition}) can be written as 
\begin{equation}
\label{partition 2}
      Z \equals  \sum_{\bn}\prod_i {z_i}^{{\bf n}(i)} 
\end{equation}
and the density operator (\ref{Gibbs})  as 
\begin{equation}
\label{Gibbs 2}
      \Gamma \ = \  \frac{1}{Z}\sum_{\bn} \prod_i {z_i}^{\bn(i)}  |\bn \rangle\!\langle \bn |   
\end{equation}
Note that each $z_i < 1$ in the boson case, as we are assuming that $Z_{\BB}$, the partition function for bosons, is finite.

The partition function can be factored:
\begin{eqnarray}
  Z_{\FF} & = & \sum_{\bn \in \FF}\prod_i {z_i}^{{\bf n}(i)}  \equals  \prod_i (1+z_i)          \label{factored partition function fermions} \\
  Z_{\BB} & = & \sum_{\bn \in \BB}\prod_i {z_i}^{{\bf n}(i)}  \equals  \prod_i (1-z_i)^{-1}   \label{factored partition function bosons} 
\end{eqnarray}
We are assuming that $  Z_{\FF}$ and $  Z_{\BB} $ are finite, which is the case if and only if $\sum z_i < \infty$.  
Substituting the factored form for $Z$ in formula (\ref{Gibbs 2}), we obtain 
\begin{equation}
\label{Gibbs fermions}
      \Gamma \ = \   \prod_j \frac{1}{1+z_j} \sum_{\bn \in \FF} \prod_i {z_i}^{\bn(i)}  |\bn \rangle\!\langle \bn |   
\end{equation}
for fermions, and 
\begin{equation}
\label{Gibbs bosons}
      \Gamma \ = \   \prod_j(1-z_j) \sum_{\bn \in \BB} \prod_i {z_i}^{\bn(i)}  |\bn \rangle\!\langle \bn |   
\end{equation}
for bosons.  In the fermion case, the parameters $z_i$ are positive numbers; in the boson case, the parameters satisfy $0 < z_i < 1$.  In both cases, $\sum z_i < \infty$.

``Free states" are those that can be represented by density operators of the form (\ref{Gibbs fermions}) or (\ref{Gibbs bosons}) such that $\sum z_i < \infty$.  Actually, we wish to generalize the form slightly, allowing some $z_i$ to equal $0$, so that some modes or orbitals may be unoccupied, and, in the fermion case, allowing some $z_i$ to equal $\infty$, so that some orbitals may be fully occupied.  To facilitate this generalization in the fermion case, we will change the parameters in (\ref{Gibbs fermions}) from $z_i$ to $p_i = z_i/(1+z_i)$, and admit the boundary cases where some $p_i = 0$ or $1$.  

We have arrived at our definitions of free density operators:

\begin{definition} 
\label{free bosons definition}
A density operator $\Gamma$ on the boson Fock space is called ``free" when it can be written as  
\begin{equation}
\label{Quasifree bosons}
   \Gamma  \equals  \sum_{\bn \in \BB} \Big\{  \prod_i (1-z_i) z_i^{\bn(i)}\Big\} |\bn \rangle\!\langle \bn |
\end{equation}
with parameter values $z_i \in [0,1)$ such that $\sum z_i < \infty$.
\end{definition}

\begin{definition} 
\label{free fermions definition}
A density operator $\Gamma$ on the fermion Fock space is called ``free" when it can be written as  
\begin{equation}
\label{Quasifree fermions}
   \Gamma \equals \sum_{\bn \in \FF}   \Big\{ \prod_i  p_i^{\bn(i)}  (1-p_j)^{1-\bn(i)}  \Big\}|\bn \rangle\!\langle \bn |
\end{equation}
with parameter values $p_i \in [0,1]$ such that $\sum p_i < \infty$.
\end{definition}
 
Note that Slater determinant states are free states according to Definition~\ref{free fermions definition}. If $\Psi$ is a normalized $n$-fermion Slater determinant wave function, 
the density operator $|\Psi\rangle\!\langle \Psi |$ can be written in the form (\ref{Quasifree fermions}) with exactly $n$ of the parameter values $p_i$ equal to $n$ and the rest equal to $0$.

\subsection{Density operators on Fock space and their $1$-matrices}
\label{$1$-matrix section}

Recall that $\DD$ denotes the class of many-particle density matrices that commute with the number operator and represent states of finite average particle number.  

Every density operator $\Delta \in \DD$ has a $1$-matrix (often called the $1$-particle density matrix or 1PDM) that we shall denote by $\gamma_{\Delta}$.   The $1$-matrix of $\Delta$ can be characterized as follows.  It is the unique Hermitian operator $\gamma_{\Delta}$ on the $1$-particle Hilbert space $\HH$ such that, for any unit vector $h \in \HH$, the matrix element $\langle h |\gamma_{\Delta} | h \rangle$ equals the average occupation of the mode $h$ when the many-particle system is in the state with density operator $\Delta$.  For example, if the Fock space vectors $|\bn\rangle$ are defined relative to an ordered orthonormal basis $(h_1,h_2,\ldots)$ of $\HH$, then 
\begin{equation}
\label{property of $1$-matrix}
     \langle h_i|\gamma_{\Delta}| h_i  \rangle   \ = \ \sum\limits_{\bn: \bn(i) = 1 }\langle \bn | \Delta |\bn\rangle   \ .
\end{equation}
The $1$-matrix $\gamma_{\Delta}$ is characterized by the property that (\ref{property of $1$-matrix}) holds for {\it all} ordered orthonormal bases of $\HH$.  

The eigenvectors of $\gamma_{\Delta}$ are called ``natural" modes or orbitals of $\Delta$, and the corresponding eigenvalues are the ``natural occupation numbers" of $\Delta$.  
If $(g_1,g_2,\ldots)$ is an orthonormal system of vectors in $\HH$, and if 
\begin{equation}
\label{gamma sub Delta}
    \gamma_{\Delta} \ = \ \sum_i n_i |g_i\rangle\!\langle g_i| \ ,
\end{equation}
so that each $g_i$ is an eigenvector of $\gamma_{\Delta}$ with eigenvalue $n_i$, then $n_i$ is indeed the occupation of $g_i$ by property (\ref{property of $1$-matrix}) of the $1$-matrix.  

Consider the free density operators $\Gamma$ of  (\ref{Quasifree bosons}) and (\ref{Quasifree fermions}), and suppose that the Fock basis used in those formulas is defined relative to an ordered orthonormal basis $(f_1,f_2,\ldots)$ of $\HH$.  These $f_i$ are in fact the natural orbitals or modes of $\Gamma$.  The corresponding natural occupation numbers are $p_i$ in the fermion case, and $z_i/(1-z_i)$ in the boson case.  That is,
\begin{equation}
\label{$1$-matrix free bosons}
      \gamma_{\Gamma} \ = \    \sum_i \frac{z_i}{1-z_i} |f_i\rangle\!\langle f_i| 
\end{equation}
if $\Gamma$ is the free boson density matrix of (\ref{Quasifree bosons}), but 
\begin{equation}
\label{$1$-matrix free fermions}
      \gamma_{\Gamma} \ = \    \sum_i p_i |f_i\rangle\!\langle f_i| 
\end{equation}
if $\Gamma$ is the free fermion density matrix of (\ref{Quasifree fermions}).

The correspondence $\Delta \mapsto \gamma_{\Delta}$ between density operators on the many-particle Fock space and their $1$-matrices is many-to-one.  That is, except in special cases, there are infinitely many density operators besides $\Delta$ that have $1$-matrix $\gamma_{\Delta}$.  However, there is only one {\it free} density operator with $1$-matrix $\gamma_{\Delta}$.  
We shall denote the unique free density operator with $1$-matrix $\gamma_{\Delta}$ by $\Gamma_{\Delta}$.

The free state $\Gamma_{\Delta}$ depends on $\Delta$ only through its $1$-matrix $\gamma_{\Delta}$.  One can see how by comparing (\ref{$1$-matrix free bosons}) or (\ref{$1$-matrix free fermions}) to (\ref{gamma sub Delta}).
Suppose $\Delta \in \DD$ and let $\gamma_{\Delta}$  be as in (\ref{gamma sub Delta}), so that $\Delta$ has natural modes $g_i$ and natural occupation numbers $n_i$.    The natural modes $f_i$ of $\Gamma_{\Delta}$ must be the same as the natural modes $g_i$ of $\Delta$.  Using the $g_i$ as the reference modes for the Fock basis $\big\{ |\bn\rangle :  \bn \in \FF \big\}$ or $\big\{ |\bn\rangle: \bn \in \BB \big\}$, the free density operator $\Gamma_{\Delta}$ is described by formula (\ref{Quasifree bosons}) or (\ref{Quasifree fermions}), wherein the parameters are related to the $n_i$ by $p_i = n_i$ in the fermion case and $z_i = n_i/(1+n_i)$ in the boson case.

\subsection{Nonfreeness and relative entropy}
\label{nonfreeness definition section}

For every $\Delta \in \DD$, there exists a unique free density operator $\Gamma_{\Delta}$ that has the same $1$-matrix as $\Delta$.   
In \cite{GottliebMauser2005}, we proposed that the ``correlation"  in a many-fermion state $\Delta$ could be quantified by comparing it to the reference state $\Gamma_{\Delta}$; the more $\Delta$ resembles $\Gamma_{\Delta}$, the less correlation it contains.  In \cite{GottliebMauser2007}, we considered the benefits of using the relative entropy $S(\Delta | \Gamma_{\Delta})$ to compare $\Delta$ to  $\Gamma_{\Delta}$.  
The relative entropy $S(\Delta | \Gamma_{\Delta})$ is non-negative, though it may equal $\infty$, and $S(\Delta | \Gamma_{\Delta}) = 0$ if and only if $\Delta = \Gamma_{\Delta}$, i.e., if and only if $\Delta$ is free.

\begin{definition} 
\label{nonfreeness definition}
Let $\Delta \in \DD$ be a density operator on the fermion or boson Fock space.  The ``nonfreeness" of $\Delta$ is defined to be $S(\Delta | \Gamma_{\Delta})$, the entropy of $\Delta$ relative to $\Gamma_{\Delta}$.  
\end{definition}

When $ \Tr ( \Delta \log \Delta ) > -\infty$,
\begin{equation}
\label{relative entropy}
      S(\Delta | \Gamma_{\Delta}) \ = \  \Tr ( \Delta \log \Delta ) -  \Tr ( \Delta \log \Gamma_{\Delta} )
\end{equation}
by definition, and then $S(\Delta | \Gamma_{\Delta}) = \infty$ if and only if $\Tr ( \Delta \log \Gamma_{\Delta} ) = -\infty$.  
In case $ \Tr ( \Delta \log \Delta ) = -\infty$, formula (\ref{relative entropy}) cannot serve to define $S(\Delta | \Gamma_{\Delta})$, which may be finite or infinite;  
the proper definition of relative entropy for such cases can be found in \cite{Lindblad73,Lindblad74}.  In Propositions~\ref{superproposition fermions} and \ref{superproposition bosons} below, we explicitly assume that $ \Tr ( \Delta \log \Delta ) > -\infty$, so that we may use formula (\ref{relative entropy}) as a working definition of relative entropy.

The monotonicity property of nonfreeness mentioned in the introduction is easily established using the mononoticity property of quantum relative entropy \cite{Lindblad75,Uhlmann77,OhyaPetz,FrankLieb}.   We will not prove it here, but refer the reader to Prop.~2 of \cite{GottliebMauser2007}.

\section{Nonfreeness as a relative entropy minimizer}
\label{the proof section}

In this section, we prove our main result, i.e., that nonfreeness is the minimum of relative entropy relative to all free reference states.  Separate propositons are stated for the two cases, fermions and bosons, though the proofs are very similar.  Because the proofs are so similar, the proof of Prop.~\ref{superproposition bosons} is abridged, and refers the reader to the proof of Prop.~\ref{superproposition fermions}.

\begin{proposition} 
\label{superproposition fermions}
Let $\Delta \in \DD$ be a density operator on the fermion Fock space and let $\Gamma_{\Delta}$ denote the unique free state that has the same $1$-matrix as $\Delta$.

Suppose that $\Tr(\Delta\log \Delta) > - \infty$.  
Then 
\begin{eqnarray}
  \lefteqn{  S(\Delta|\Gamma_{\Delta})  \equals  - \Tr(\Delta\log \Gamma_{\Delta}) \ + \ \Tr(\Delta\log \Delta) } \nonumber \\
    & = &  - \sum_i  n_i\log(n_i) - \sum_i  (1-n_i)\log(1-n_i) \ + \ \Tr(\Delta\log \Delta) 
\label{first assertion fermions}
\end{eqnarray}   
and $S(\Delta|\Gamma_{\Delta})$ is the minimum value of $S(\Delta|\Gamma _{free})$ over all free density matrices, that is, 
\begin{equation}
    S(\Delta|\Gamma_{\Delta}) \ = \ \min \big\{S(\Delta|\Gamma _{free}) : \ \Gamma _{free} \ \mathrm{is}\ \mathrm{ free} \big\}.
\label{second assertion fermions}
\end{equation}   
If $\Tr(\Delta\log \Gamma_{\Delta})  > - \infty$, then the minimum in (\ref{second assertion fermions}) 
is attained only at $\Gamma_{\Delta}$.
\end{proposition}

\noindent{\bf Proof:}  \qquad  

Let $\Gamma$ be any free density matrix on the fermion Fock space, as in (\ref{Quasifree fermions}).   
Then
\begin{eqnarray*}
\lefteqn {\log \Gamma  
      \equals
        \sum_{\bn \in \FF}  \sum_i  \Big(   \bn(i) \log(p_i) + (1-\bn(i)) \log(1-p_i)   \Big) \ |  \bn \rangle\!\langle\bn  | } \\
       & = & 
       \sum_i   \log(p_i) \sum_{\bn: \bn(i) = 1 } | \bn  \rangle \! \langle\bn  |  \ + \  \sum_i   \log(1-p_i) \sum_{\bn: \bn(i) =  0 } | \bn  \rangle\!\langle\bn  |  \ ,
\end{eqnarray*}
and therefore $\Tr(\Delta \log\Gamma )$ equals 
\begin{eqnarray}
\lefteqn{ \sum_i  \Big[ \log(p_i) \sum_{\bn: \bn(i) = 1 }\Tr\big( \Delta |\bn  \rangle\!\langle\bn  | \big) 
   + 
      \log(1-p_i) \sum_{\bn: \bn(i) =  0 } \Tr\big( \Delta |\bn  \rangle\!\langle\bn  | \big) \Big] } \nonumber \\
   & = &
   \sum_i \Big[  \log(p_i) \sum_{\bn: \bn(i) = 1 }\langle \bn | \Delta |\bn\rangle 
    +
          \log(1-p_i) \sum_{\bn: \bn(i) =  0 } \langle \bn | \Delta |\bn\rangle \Big] \ . 
\label{maximizeme1}
\end{eqnarray}
Now $\sum\limits_{\bn: \bn(i) = 1 }\langle \bn | \Delta |\bn\rangle$ is equal to $\langle f_i|\gamma_{\Delta}| f_i  \rangle$, 
for both express the occupation probability of the mode $f_i$ when the system is in the state $\Delta$, as in formula (\ref{property of $1$-matrix}).
Thus  
\begin{equation}
  \Tr(\Delta \log\Gamma ) 
  \ = \ \sum_i \Big[  \log(p_i)\langle f_i|\gamma_{\Delta}| f_i  \rangle 
    +
          \log(1-p_i) \big(1 - \langle f_i|\gamma_{\Delta} | f_i  \rangle\big) \Big]   \ .  
\label{maximizeme2 fermions}
\end{equation}

Let $\Phi$ denote the strictly convex function on the domain $[0,1]$ defined by $\Phi(0)=\Phi(1)=0$ and 
\begin{equation}
\Phi (x) \ = \   x\log(x) + (1-x)\log(1-x)
\label{definition of Phi fermions}
\end{equation}
for $0 < x < 1$.  
Since 
$$ x\log(p) + (1-x) \log(1-p) \ \le \ \Phi(x)$$ 
whenever $0 \le p,x \le 1$, it follows from  (\ref{maximizeme2 fermions}) that 
\begin{equation}
   \Tr (\Delta \log \Gamma )  \ \le \ \sum_i \ \Phi \big(\langle f_i|\gamma_{\Delta} f_i  \rangle\big)\ .
\label{maximizeme3 fermions}
\end{equation}
In case $\Gamma = \Gamma_{\Delta}$, 
the parameters $p_i$ are natural occupation numbers of $\Delta$  
and the vectors $f_i$ are natural orbitals of $\Delta$, as can be seen by comparing 
(\ref{gamma sub Delta}) and (\ref{$1$-matrix free fermions}).  In this case, it follows from 
(\ref{maximizeme2 fermions}) that
\begin{equation}
  \Tr(\Delta \log\Gamma_{\Delta}) 
  \ = \ \sum_i\Phi(n_i)    \ ,
\label{recall me fermions}
\end{equation}
where the $n_i$ are the natural occupation numbers of $\Delta$.

If general, the $f_i$ are not natural orbitals of $\Delta$, that is, not eigenvectors of $\gamma_{\Delta}$. 
Let $g_1,g_2,\ldots$ be an ordered orthonormal basis of $\HH$ consisting of eigenvectors of $\gamma_{\Delta}$, and denote the corresponding eigenvalues by $n_1,n_2,\ldots$, 
so that $\gamma_{\Delta}g_i = n_ig_i$
for all $i$.  
The orthonormal bases $f_1,f_2,\ldots$ and $g_1,g_2,\ldots$ are related by a unitary transformation with matrix elements $u_{ij} = \langle g_j | f_i\rangle$, so that $f_i = \sum_j u_{ij}g_j$ for all $i$.
Since $g_1,g_2,\ldots$ is an ``eigen-basis" of $\gamma_{\Delta}$, 
$$ 
\langle f_i|\gamma_{\Delta} f_i  \rangle  =  \Big\langle \sum_k u_{ik}g_k \big| \ \sum_j u_{ij} n_j g_j   \Big\rangle  = \sum_j |u_{ij}|^2 n_j \ .
$$
Since $\sum_j |u_{ij}|^2=1$ and $\Phi$ is convex, Jensen's inequality implies that 
\begin{equation}
\label{Jensen implies}
\Phi \big(\langle f_i|\gamma_{\Delta} f_i  \rangle\big) \ \le\   \sum_j |u_{ij}|^2\Phi( n_j ) \ .
\end{equation}
Substituting this in (\ref{maximizeme3 fermions}) and using the fact that $\sum_i |u_{ij}|^2$ is also equal to $1$, we find that 
\begin{eqnarray}
\Tr(\Delta \log\Gamma ) 
 & \le & 
   \sum_i \Phi \big(\langle f_i|\gamma_{\Delta} f_i  \rangle\big)   \label{then look here} \\
   & \le & \sum_i  \sum_j |u_{ij}|^2\Phi( n_j )      \label{look here}  \\
   & = &   \sum_j  \sum_i |u_{ij}|^2\Phi( n_j )  \equals  \sum_j \Phi( n_j )    \nonumber \\
   & \stackrel{(\ref{recall me fermions})}{=}  & 
       \Tr(\Delta \log\Gamma_{\Delta}) \nonumber
\end{eqnarray}
(changing the order of summation is justified even when $\Tr(\Delta \log\Gamma ) = - \infty$ because every term in the double series (\ref{look here}) is $\le0$).
We have now shown that 
\begin{eqnarray}
 \Tr(\Delta \log\Gamma ) \ \le\  \Tr(\Delta \log\Gamma_{\Delta})
\label{we've now shown this}
\end{eqnarray} 
for an arbitrary free density matrix $\Gamma$.  Therefore, 
\begin{eqnarray*}
S(\Delta|\Gamma_{\Delta}) & = &   \Tr(\Delta\log\Delta) \ - \ \Tr(\Delta \log\Gamma_{\Delta})   \\
& \le &  \Tr(\Delta\log\Delta) \ - \ \Tr(\Delta \log\Gamma ) \ = \ S(\Delta|\Gamma )
\end{eqnarray*}
holds for every free density matrix $\Gamma $.  This proves (\ref{second assertion fermions}).   

Assertion (\ref{first assertion fermions}) follows from formulas (\ref{relative entropy}) and (\ref{recall me fermions}).

Finally, suppose that $\Tr(\Delta\log \Gamma_{\Delta})  > - \infty$ and let $\Gamma$ be a free density matrix such that 
$S(\Delta|\Gamma)=S(\Delta|\Gamma_{\Delta})$.  We will prove that $\Gamma=\Gamma_{\Delta}$.  

Since $S(\Delta|\Gamma)=S(\Delta|\Gamma_{\Delta})$, there must be equality at (\ref{look here}).  Because of our assumption that $\Tr(\Delta\log \Gamma_{\Delta})  > - \infty$, equality can hold at (\ref{look here}) only if it holds in each application (\ref{Jensen implies}) of Jensen's inequality.   Since $\Phi$ is strictly convex, this implies that for each $i$ there exists $i'$ such that $u_{ij}=0$ unless $n_j = n_{i'}$.  Since $f_i = \sum_j u_{ij}g_j$,  equality holds in (\ref{Jensen implies}) only if $f_i$ is in the eigenspace of $\gamma_{\Delta}$ for eigenvalue $n_{i'}$.  Thus the orthonormal basis $(f_i)$ must be an eigen-basis of $\gamma_{\Delta}$.  

Again, since $S(\Delta|\Gamma)=S(\Delta|\Gamma_{\Delta})$, there must also be equality at (\ref{then look here}).  Looking back to the argument from (\ref{maximizeme2 fermions}) to (\ref{maximizeme3 fermions}), we see that this can happen only if $p_i = \langle f_i|\gamma_{\Delta} f_i \rangle$ for all $i$.  Since $f_i$ is an eigenvector of  $\gamma_{\Delta}$, $p_i$ is  the corresponding eigenvalue.  Therefore, the $1$-matrix (\ref{$1$-matrix free fermions}) of $\Gamma$ must equal $\gamma_{\Delta}$, and $\Gamma$ must equal $\Gamma_{\Delta}$.
\hfill $\square$


\begin{proposition} 
\label{superproposition bosons}
Let $\Delta \in \DD$ be a density operator on the boson Fock space and let $\Gamma_{\Delta}$ denote the unique free state that has the same $1$-matrix as $\Delta$.

Suppose that $\Tr(\Delta\log \Delta) > - \infty$. 
Then 
\begin{equation}
    S(\Delta|\Gamma_{\Delta})   \ = \    \sum_i  (1+n_i)\log(1+n_i) - \sum_i  n_i\log(n_i) \ +\ \Tr(\Delta\log \Delta)
\label{first assertion bosons}
\end{equation}   
and $S(\Delta|\Gamma_{\Delta})$ is the minimum value of $S(\Delta|\Gamma _{free})$ over all free density matrices, that is, 
\begin{equation}
    S(\Delta|\Gamma_{\Delta}) \ = \ \min \big\{S(\Delta|\Gamma _{free}) : \ \Gamma _{free} \ \mathrm{is}\ \mathrm{ free} \big\}.
\label{second assertion bosons}
\end{equation}   
If $\Tr(\Delta\log \Gamma_{\Delta})  > - \infty$, then the minimum in (\ref{second assertion bosons}) 
is attained only at $\Gamma_{\Delta}$.
\end{proposition}

\noindent{\bf Proof:}  \qquad  

Let $\Gamma$ be any free density matrix on the boson Fock space, as in (\ref{Quasifree bosons}).    
Then
\begin{equation}
   \log \Gamma \equals 
   \sum_{\bn \in \BB} \Big\{  \sum_i \log(1-z_i) + \bn(i) \log( z_i)\Big\} |\bn \rangle\!\langle \bn |
 \label{logQuasifree}
\end{equation}
and therefore 
\begin{eqnarray}
 \Tr(\Delta \log\Gamma)  & = & 
 \sum_i \log(1-z_i)  + \sum_{\bn \in \BB} \sum_i  \bn(i) \log( z_i)   \langle \bn |\Delta |\bn \rangle \nonumber \\
   & = & 
 \sum_i \log(1-z_i)  + \sum_i \Big\{ \log( z_i) \sum_{\bn}  \bn(i)   \langle \bn |\Delta |\bn \rangle \Big\}  \nonumber \\
   & = &
    \sum_i \log(1-z_i)  +  \sum_i   \log( z_i)   \langle f_i |\gamma_{\Delta} | f_i \rangle   \ . 
\label{maximizeme2 bosons}
\end{eqnarray}

Let $\Phi (x)$ be the strictly convex function defined on $[0,\infty)$ by $\Phi(0)=0$ and
\begin{equation}
  \Phi (x)\ = \  x\log(x) - (1+x)\log(1+x)
\label{definition of Phi bosons}
\end{equation}
when $x>0$.
The $i$'th term of the series (\ref{maximizeme2 bosons}) is maximized when $$z_i =  \frac{ \langle f_i | \gamma_{\Delta} f_i\rangle }{  1+  \langle f_i | \gamma_{\Delta} f_i\rangle }\ ,  $$
so that 
\begin{equation}
   \Tr (\Delta \log \Gamma)  \ \le \ \sum_i \ \Phi  \big(\langle f_i|\gamma_{\Delta} f_i  \rangle\big)\ .
\label{maximizeme3 bosons}
\end{equation}

In the case where $\Gamma = \Gamma_{\Delta}$, the parameters $z_i$ in (\ref{Quasifree bosons}) are related to the state's natural occupation numbers $n_i$ as $n_i = z_i/(1-z_i)$, 
 the vectors $f_i$ are the natural modes of $\Delta$, and $n_i = \langle f_i|\gamma_{\Delta} f_i  \rangle$.   In this case, it follows from 
(\ref{maximizeme2 bosons}) that
\begin{equation}
  \Tr(\Delta \log\Gamma_{\Delta}) 
  \ = \ \sum_i\Phi(n_i)    \ .
\label{recall me bosons}
\end{equation}
Assertion (\ref{first assertion bosons}) now follows from formulas (\ref{recall me bosons}) and (\ref{definition of Phi bosons}).

Assertion (\ref{second assertion bosons}) and the uniqueness of the minimizer may be proved exactly as was done in the proof of Proposition~\ref{superproposition fermions}, using Jensen's inequality.   \hfill $\square$

\section{Examples of nearly free substates}
\label{examples section}

In this section we consider two examples, both of which concern essentially non-interacting systems.   The first example is about pure states of $n$ fermions in $m$ orbitals, random pure states whose wave functions are sampled uniformly from the unit sphere in the system's $\binom{m}{n}$-dimensional Hilbert space.  The second example is the $n$-particle canonical ensemble for the ideal Bose gas.  

The states we consider are not free, but not because of interactions between the particles.  It is only the restriction on total number $n$ that keeps them from being free.  However, as $n$ increases, this global constraint has less and less effect on the state of a low-dimensional subsystem.  The nonfreeness of the substate decreases to $0$ even as the nonfreeness of the global state increases without bound.

\subsection{Random substates}
\label{random substates}

Consider a pure state $\Psi$ of $n$ fermions in $m \ge n$ orbitals, and the derived state delimited by  $s$ of those orbitals.  If the $n$-fermion pure state $\Psi$ is sampled randomly according to surface area on the unit sphere in the $n$-fermion Hilbert space, and if $s \ll m$, then the substate of the fermions in the first $s$ orbitals is expected to be nearly free.  

For example,  we generated $10^6$ pure states of $n=4$ fermions in $m=8$ orbitals by sampling the wave vector pseudo-randomly from the unit sphere in $2\binom{8}{4}=140$-dimensional real space.   From each pure state, we derived the substate of the fermions in the first $s=2$ orbitals.    The average  nonfreeness of the substate was just $0.022$, much smaller than the average von Neumann entropy of the substate, which was $1.93$ bits (we use base-2 logarithms).

Here is a little table that shows typical values of the nonfreeness of the state of the first two orbitals, for various values of $n$ and $m$.  $10^6$ pseudo-random trials were done for each $(n,m)$, except for $(n,m) = (8,16)$, where only $10^4$ trials were done.  
To provide a sense of scale, the median of the substates' von Neumann entropies is tabulated alongside the quartiles of nonfreeness.

\begin{center}
\begin{tabular}{ c c | c  |   ccc }
   &    &   
    Median   & 
    $1^{st}$ quartile  & Median  &  $3^{rd}$ quartile  \\
 $n$ &  $m$  & 
 entropy  & 
 nonfreeness  & nonfreeness & nonfreeness   \\
 \hline
  4 & 8    
  & 1.942   
  & 0.003 & 0.013 & 0.032 \\ 
  5 & 10  
  & 1.979   
  & 0.003 & 0.008 & 0.016 \\
  6 & 12  
  & 1.991  
  & 0.003 & 0.006 & 0.009 \\
  7 & 14  
  & 1.995   
  & 0.003 & 0.004 & 0.006 \\
  8 & 16  
  & 1.997   
  & 0.003 & 0.003 & 0.004 \\
  \hline
   6 & 8 
   & 1.488  
   & 0.005 & 0.018  & 0.044 \\
   9 & 12 
   & 1.601 
   &  0.002 & 0.006 &  0.013 \\
 12 & 16 
  & 1.617   
  &   0.002 & 0.003 & 0.005    \\
\end{tabular}
\end{center}
\begin{center}
Table I
\end{center}
\medskip

Observe that the nonfreeness of the substate is rather small, compared to the von Neumann entropy.   This is due to the ``concentration of measure" effect, whereby  ``almost all quantum states behave in essentially the same way" when they are sampled  uniformly from the unit sphere  \cite{HaydenLeungWinter}.   

The concentration of measure effect is seen on the reduced density matrices themselves, as explained in \cite{PopescuShortWinter-arxiv,PopescuShortWinter-Nature}.  
We review the general result of \cite{PopescuShortWinter-Nature} before applying it to our specific setting:  

Let $\HH_S$ and $\HH_E $ denote the Hilbert spaces for a system $S$ of interest and its ``environment" $E$, so that $\HH_S \otimes \HH_E $ is the Hilbert space for the system-plus-environment.  Assume these Hilbert spaces are finite dimensional.  Suppose we have a restriction on the states  in $\HH_S \otimes \HH_E $, e.g., a constraint on energy and/or other conserved quantities, such as particle number.  Let $\HH_R$ denote the subspace of $\HH_S \otimes \HH_E $ consisting of wave functions 
that satisfy that restriction, and let $\mathcal{E}_R$ denote the density matrix that is proportional to the orthogonal projector onto $\HH_R$.  This is the density matrix that represents the ``microcanonical" ensemble of the system-plus-environment with restriction $R$.  
The reduced density matrix for the state of the system $S$, a substate of $\mathcal{E}_R$, is 
\[
    \Omega_S \ = \ \Tr_E \mathcal{E}_R\ .
\]
The mixed state $\mathcal{E}_R$ can be realized by sampling random wave functions $\phi$ uniformly from the unit sphere in $\HH_R$.   When $\dim\HH_R$ is much larger than $\dim\HH_S$,   
concentration of measure for high-dimensional spheres implies that  the random substate $\Tr_E |\phi\rangle\!\langle \phi|$ is very close to the average substate $\Omega_S$ with high probability  \cite{HaydenLeungWinter,PopescuShortWinter-arxiv,PopescuShortWinter-Nature}.  

In many physical models, where a small subsystem is weakly coupled to the environment, the substate $\Omega_S$ tends toward a canonical equilibrium state in the thermodynamic limit.  
When this is the case, the concentration of measure phenomenon leads to the ``canonical typicality" observed in \cite{CanonicalTypicality}, that is, the fact that ``in the thermodynamic limit, the reduced density matrices of the overwhelming majority of the wave functions of [the system-plus-environment] are canonical."  

In our setting, there are $n$ fermions in $m$ orbitals, the system $S$ is the subsystem consisting of $s$ orbitals and the particles that occupy them, and the environment $E$ is the subsystem delimited by the rest of the orbitals. 
The fermion Fock space over all of the orbitals is the tensor product  $\HH_S \otimes \HH_E $, but we are exclusively interested in the subspace $\HH_R$ of  that Fock space consisting only of $n$-fermion wave functions.  
The substate $\Omega_S$ of the microcanonical equilibrium state $\mathcal{E}_R$ is free in the thermodynamic limit.  That is, if $s$ is fixed and $n,m$ tend to infinity with constant ratio $\rho = n/m$, then $\Omega_S$ tends to a free density matrix on the Fock space $\HH_S$ of the subsystem.  This free density matrix is the one in which the natural oribitals are occupied independently with probability $\rho$.   

``Canonical typicality" explains why our random substates have low nonfreeness.   
It is because the random substates are close to $\Omega_S$ when $s \ll \binom{m}{n}$, and $\Omega_S$ is nearly free when $m$ and $n$ are large enough.




\subsection{Ideal canonical bosons}
\label{ideal canonical bosons}

Consider the canonical thermal equilibrium state for a system of $n$ non-interacting bosons that may occupy the ground mode of energy $\epsilon_0 = 0$ or the ``excited" modes of energies $\epsilon_1, \epsilon_2, \ldots > 0$.      
There may be infinitely many modes, but we assume that $\sum_i e^{-\beta\epsilon_i}$ is finite,  where $\beta = 1/k_BT$.  
As in Section~\ref{free states section}, let us denote $n$-boson configurations by their occupation lists
\[
     \bn = \big(\bn(0),\bn(1),\bn(2),\bn(3),\ldots \big)
\]
relative to these modes, where the mode with index $0$ is the ground mode.  The density operator for the canonical thermal equilibrium state is 
\begin{equation}
\label{canonical thermal}
     \frac{1}{Z_n}   \sum_{\bn \in \BB_n} e^{-\beta \sum_i \bn(i) \epsilon_i } |\bn \rangle\!\langle \bn |   \ ,
\end{equation}
where $\BB_n = \big\{ \bn \in \BB: \sum \bn(i) = n\big\}$ and 
\begin{equation}
\label{canonical partition}
   Z_n \equals  \sum_{\bn \in \BB_n} e^{-\beta \sum_i \bn(i) \epsilon_i }\ .
\end{equation}
Fix $\beta >0$ and consider the limit $n \longrightarrow \infty$.  As $n$ increases, most of the bosons pile into the ground mode, but a residual number of them populate the excited modes.  In the limit, the state of the bosons in the excited modes is described by the {\it grand} canonical ensemble for temperature $T$ and chemical potential $\mu = 0$ \cite{Politzer,HolthausKalinowskiKirsten}.

Let us look at a simple case: a system of $n$ bosons occupying $3$ modes, a ground mode $f_0$ and two excited modes $f_1$ and $f_2$.   Consider a state $\Delta$ of the three-mode system that has the canonical thermal density operator  (\ref{canonical thermal}).    
Let $\Delta_0$ denote the state of the the ground mode $f_0$ and its excitations, a substate of $\Delta$.  Similarly, let $\Delta_{12}$ denote the state of modes $f_1$ and $f_2$ and their excitations when the entire system is in the state   $\Delta$.    As $n$ increases, the bosons pile in the ground mode, and the substate $\Delta_{12}$ tends toward freeness.  

The following table shows the nonfreeness of $\Delta$ and its substates, for several values of $n$ and parameters $x_1 =  e^{-\beta \epsilon_1 }$ and $x_2 =  e^{-\beta \epsilon_2 }$.  In this table, nonfreeness has been rounded to the third decimal place; entries $0.000$ are not exactly $0$.   In Appendix~2 we explain how the nonfreeness of $\Delta $, $\Delta_0$,  and $\Delta_{12}$  was computed.

\begin{center}
\begin{tabular}{ c c c   | c c c  c c }
   $x_1$  & $x_2$ & $n$ &  $\Delta $ & $\Delta_0$ &  $\Delta_1$ &  $\Delta_2$ &  $\Delta_{12}$   \\
   \hline
 $\tfrac12$ & $\tfrac12$ &  4 &   3.143  & 0.823 &   0.026 &   0.028 &   0.109  \\
                    &                   & 8 &   4.175  & 1.531 &   0.003 &   0.003 &   0.014   \\
                    &                   & 12 &   4.837&  2.135 &   0.000 &  0.000 &  0.001  \\
 \hline
 $\tfrac23$ & $\tfrac12$ & 8 &  4.064 & 1.092 &   0.027 &   0.003 &   0.056  \\
                    &                   & 12 &   4.722  &1.591 &  0.007 & 0.000 & 0.013  \\
                    &                   &  16 &    5.204  & 2.025 & 0.002 &    0.000 &    0.003  \\
  \hline
 $\tfrac45$ & $\tfrac23$ & 12 &   4.510 &   0.832 &   0.047 &    0.006 &   0.105 \\
                    &                   &  16 &   4.977 & 1.114 &   0.025 &   0.002 &   0.050  \\
                    &                   &  20 &   5.358  & 1.396 &   0.002 &   0.000 &    0.002  \\
\end{tabular}
\end{center}
\begin{center}
Table II
\end{center}
\medskip

Observe that the  nonfreeness of $\Delta_{12} $ decreases toward $0$ as $n$ increases, while the nonfreeness of the ground mode substate $\Delta_0$ increases.   Note that the substates $\Delta_1$ and  $\Delta_2$ have even lower nonfreeness than $\Delta_{12}$, as implied by the monotonicity of nonfreeness.

\section{Conclusion}
\label{conclusion}

We have considered an independent-particle model 
in which the ``free" states epitomize independent-particle behavior. 
Free states are those that  have the form of a Gibbs grand canonical ensemble of non-interacting particles.  In the fermion case, the class of free states includes all Slater determinant states.

We have characterized the free state that minimizes the relative entropy $S(\Delta|\Gamma _{free})$ between a state of interest $\Delta$ and the various free states $\Gamma _{free}$.   
It is $\Gamma_{\Delta}$, the unique free density operator with the same $1$-matrix as $\Delta$.    
That is, 
$$ S(\Delta|\Gamma_{\Delta}) \ = \ \min \big\{S(\Delta|\Gamma _{free}) : \ \Gamma _{free} \ \mathrm{is}\ \mathrm{ free} \big\}\ . $$
In this sense, $\Gamma_{\Delta}$ is an optimal independent-particle approximation of $\Delta$. 

The relative entropy $ S(\Delta|\Gamma_{\Delta})$ is called the ``nonfreeness" of $\Delta$.     
It equals $0$ if and only if $\Delta$ is a free state, otherwise  it is positive, possibly $+\infty$.  Thanks to the amenable form of $\Gamma_{\Delta}$, the nonfreeness of $\Delta$ can be expressed simply as in (\ref{nonfreeness fermions}) and (\ref{nonfreeness bosons}).   
Nonfreeness is defined for mixed states as well as pure states, and has a desirable monotonicity property: the nonfreeness of  a substate is less than or equal to the nonfreeness of the state from which that substate is derived.   


\section{Acknowledgments}
 
 This work has been supported by the Austrian Science Foundation FWF, project SFB F41 (VICOM) and project I830-N13 (LODIQUAS). 
The results in Section~\ref{random substates} were obtained in collaboration with Rada M. Weish\"aupl, who is supported by the FWF project T402-N13.

\section {Appendix}

In this appendix we explain how we computed the nonfreeness values in Table~II of Section~\ref{ideal canonical bosons}.  
Formulas for the nonfreeness of $\Delta$, $\Delta_0$, and $\Delta_{12}$ are given in (\ref{nonfreeness of 012}), (\ref{nonfreeness of 0}), and (\ref{nonfreeness of 12}), respectively.  There are similar formulas for the nonfreeness of  $\Delta_1$ and  $\Delta_2$.

The density matrix of the canonical thermal distribution  (\ref{canonical thermal})  is
\begin{equation}
\label{example canonical thermal}
    \Delta \equals  \frac{1}{Z_n(x_1,x_2)}   \sum_{\bn = (n-k_1 -k_2,k_1,k_2)} x_1^{k_1}x_2^{k_2} |\bn \rangle\!\langle \bn |   
\end{equation}
with $  Z_n(x_1,x_2)  =   \sum\limits_{k=0}^n \sum\limits_{j=0}^k x_1^j x_2^{k-j}$.

Modes $f_0,f_1,$ and $f_2$ are indeed the natural modes of $\Delta$, and the natural occupation numbers are 
\[ 
    n_i \equals  \frac{1}{Z_n(x_1,x_2)}  \sum_{k_0,k_1,k_2:\ k_0+k_1+k_2=n} k_i \ x_1^{k_1} x_2^{k_2}
\]
for $i=0,1,2$.  
The von Neumann entropy of the density operator $\Delta$ is equal to the Shannon entropy of the probability distribution 
$$\PPP_n(k_0,k_1,k_2)  = x_1^{k_1}x_2^{k_2}/Z_n(x_1,x_2), $$ which is found to be 
$$ 
S(\Delta) \ = \    \log_2(Z_n(x_1,x_2)) - n_1 \log_2(x_1) - n_2 \log_2(x_2)\ .$$
By formula (\ref{nonfreeness bosons}), the nonfreeness of $\Delta$ is 
\begin{eqnarray}
\label{nonfreeness of 012}
  \lefteqn{ \sum_{i=0,1,2}  \big( (n_i+1)\log_2(n_i+1)- n_i\log_2(n_i) \big)  } & \nonumber \\
 &  - \log_2(Z_n(x_1,x_2)) + n_1 \log_2(x_1) + n_2 \log_2(x_2)\ .
\end{eqnarray}

The substate $\Delta_0$ is the state of the excitations of the ground mode $f_0$.  
 The von Neumann entropy of the density operator $\Delta_0$ is simply the Shannon entropy of the marginal distribution $p_0$ defined by 
$$  p_0(k) \equals  \PPP_n\{ \bn(0) = k \} \equals \frac{1}{Z_n(x_1,x_2)} \sum_{k_1,k_2:\ k_1+k_2=k }  x_1^{k_1}x_2^{k_2} $$
for $ k = 0,1,\ldots,n $ and $p_0(k)  = 0$ for $k>n$.  According to formula (\ref{nonfreeness bosons}), the nonfreeness of $\Delta_0$ is 
\begin{equation}
\label{nonfreeness of 0}
      (n_0+1)\log_2(n_0+1) - n_0\log_2(n_0)   - \sum_{k=0}^n  p_0(k) \log_2(p_0(k))\ .
\end{equation}

The substate $\Delta_{12}$ is the state of modes $f_1$ and $f_2$ and their excitations. 
The von Neumann entropy of the density operator $\Delta_{12}$ is the Shannon entropy of the marginal distribution $p_{12}$ defined by 
$$  p_{12}(k_1,k_2) \equals  \PPP_n\{ \bn(1) = k_1,  \bn(2) = k_2 \} \equals \frac{x_1^{k_1}x_2^{k_2} }{Z_n(x_1,x_2)} $$
for $ k_1+k_2 \le n $ and $p_{12}(k_1,k_2)  = 0$ when $ k_1+k_2 > n $.   The entropy of $\Delta_{12}$ is thus the same as that of $\Delta$, whence the nonfreeness of $\Delta_{12}$ equals 
\begin{eqnarray}
\label{nonfreeness of 12}
  \lefteqn{ \sum_{i=1,2}  \big( (n_i+1)\log_2(n_i+1)- n_i\log_2(n_i) \big)  } & \nonumber \\
 &  - \log_2(Z_n(x_1,x_2)) + n_1 \log_2(x_1) + n_2 \log_2(x_2)\ .
\end{eqnarray}

\end{document}